# Chip-to-chip optical multimode communication with universal mode processors

Bo Wu[1,†], Wenkai Zhang[1,†], Hailong Zhou[1,*], Jianji Dong[1,*] Dongmei Huang[2,3], P. K. A. Wai[4] and Xinliang Zhang[1]

[1] Wuhan National Laboratory for Optoelectronics, School of Optical and Electronic Information, Huazhong University of Science and Technology, Wuhan 430074, China

[2] The Hong Kong Polytechnic University Shenzhen Research Institute, Shenzhen 518057, China

[3] Photonics Research Institute, Department of Electrical Engineering, The Hong Kong Polytechnic University, Hong Kong, 999077, China

[4] Department of Physics, Hong Kong Baptist University, Kowloon Tong, Hong Kong, 999077, China

[†]These authors contributed equally to this work

[*]Corresponding author: hailongzhou@hust.edu.cn; jjdong@hust.edu.cn;

## Abstract

The increasing amount of data exchange requires higher-capacity optical communication links. Mode division multiplexing (MDM) is considered as a promising technology to support the higher data throughput. In an MDM system, the mode generator and sorter are the backbone. However, most of the current schemes lack the programmability and universality, which makes the MDM link susceptible to the mode crosstalk and environmental disturbances. In this paper, we propose an intelligent multimode optical communication link using universal mode processing (generation and sorting) chips. The mode processor consists of a programmable 4×4 Mach Zehnder interferometer (MZI) network and can be intelligently configured to generate or sort both quasi linearly polarized (LP) modes and orbital angular momentum (OAM) modes in any desired routing state. We experimentally establish a chip-to-chip MDM communication system. The mode basis can be freely switched between four LP modes and four OAM modes. We also demonstrate the multimode optical communication capability at a data rate of 25 Gbit/s. The proposed scheme shows significant advantages in terms of universality, intelligence, programmability and resistance to mode crosstalk, environmental disturbances and fabrication errors, demonstrating that the MZI-based reconfigurable mode processor chip has great potential in long-distance chip-to-chip multimode optical communication systems.

Keywords: Optical mode generation/sorting, Multimode communication, Optical matrix network, Reconfigurability

# Introduction

The emergence of big data domain is driving the pursuit of higher data transmission rate in a single optical fiber. Different physical dimensions of light, such as wavelength, polarization, and mode, have been exploited to increase the optical communication capacity [1,2]. Unlike the wavelength and polarization dimensions which have been widely employed in the real optical communication system, the multiplexing of optical modes in multimode fibers is still in the laboratory. Various mode bases such as linearly polarized (LP) modes, orbital angular momentum (OAM) modes and other optical modes in multicore fiber have been extensively studied [3,4] and their transmission media (multimode fiber, ring-core fiber, and multi-core fiber) have also been developed to support the long-distance communication [5]. However, the reported chip-to-chip MDM communication links are still in the short-distance domain (less than 10 m) [2,6], because MDM system is still facing a tough challenge of mode crosstalk, which is ubiquitous in the process of mode generation, transmission and sorting. Since the traditional mode generator and sorter lack effective programmability [7-14], the long-distance MDM communication links tend to require complex algorithms deployed on electronic digital signal processing (DSP) to compensate the mode crosstalk and environmental disturbances, which largely increases the extra compensating cost.

The Mach-Zehnder interferometer (MZI)-based optical mesh is capable of performing any linear transformation on chip with strong reconfigurability and has been applied to optical matrix computation, optical signal processing, optical polarization/mode processing, etc [15-21]. It has shown great potential in the optical multiple-input multiple-output (MIMO) descrambler because it can perform the inverse transformation of the crosstalk matrix in the multi-core-fiber-based MDM communication system [22,23]. Similarly, the MZI mesh is promising to resolve the mode crosstalk in the multimode-fiber-based communication system once the orthogonal fiber modes can be properly generated or sampled on chip.

In this paper, we show theoretically and experimentally that the reconfigurable optical unitary MZI network can act as both a mode generator and a mode sorter in an intelligent MDM communication system. The mode processor is scalable and can be intelligently switched between LP and OAM operation with any routing state and low mode crosstalk. In the experiment, a 25 GHz high-speed intelligent communication is achieved in a chip-to-chip link. Moreover, the mode processor is compatible with other multiplexing technology in a single chip. Overall, our results demonstrate that the proposed architecture is promising to handle complex application scenarios in the long-distance multimode communication.

# Results

## Principle and chip fabrication

The chip-to-chip multimode communication setup is presented in Fig. 1(a). The basic building block of the mode processor (generator and sorter) is a 4×4 MZI unitary matrix as shown in Fig. 1(b). Each MZI consists of two phase shifters on its inner and outer arms. By adjusting the phase shifters in the network, any 4×4 unitary transmission matrix can be realized (See S1 of Supplementary material). For the mode sorter, there is a 2×2 sampling grating array that can orthogonally sample the LP and OAM modes (Fig. 1(c)). For the LP and OAM modes, the orthogonal sampling matrix can be written as

$$S_{LP} = \frac{1}{2}\begin{pmatrix} 1 & 1 & 1 & 1 \\ 1 & -1 & 1 & -1 \\ 1 & -1 & -1 & 1 \\ 1 & 1 & -1 & -1 \end{pmatrix}, S_{OAM} = \frac{1}{2}\begin{pmatrix} 1 & 1 & 1 & 1 \\ 1 & i & -i & -1 \\ 1 & -1 & -1 & 1 \\ 1 & -i & i & -1 \end{pmatrix}, \quad (1)$$

where the column vectors from left to right represent the complex amplitude of the $LP_{01}$, $LP_{11a}$, $LP_{11b}$ and $LP_{21}$ modes (or $OAM_0$, $OAM_1$, $OAM_{-1}$ and $OAM_2$ modes) sampled by the 2×2 sampling grating array, respectively. Then the MZI unitary matrix routes the sampled orthogonal modes to the desired output ports. As an example, the unitary transformation for (Mode1-Port2, Mode2-Port3, Mode3-Port4, Mode4-Port1) can be written as

$$\begin{pmatrix} 0 & 0 & 0 & 1 \\ 1 & 0 & 0 & 0 \\ 0 & 1 & 0 & 0 \\ 0 & 0 & 1 & 0 \end{pmatrix} = U_{MZI} S_{LP/OAM}, \quad (2)$$

where $U_{MZI}$ is the transmission matrix of a 4×4 MZI mesh. Owing to the orthogonality of the input and output matrices, there is always a unitary matrix $U_{MZI}$ that can satisfy any possible routing. The mode generator can be seen as a reverse processor of the mode sorter. In this case, the 2×2 grating array serves as a mode emitting antenna. Although the generated mode is a sampled version of the original mode, it can only excite the corresponding original mode in the multimode fiber owing to the orthogonality. For ease of discussion, the quasi-LP and quasi-OAM modes generated by the chip will be referred to simply as LP and OAM modes in the remainder of this paper. By combining the mode generator and the mode sorter, a complete chip-to-chip MDM system can be constructed (Fig. 1(a)). The free space is used to transmit the optical modes, while a multimode fiber can also be used for long-distance communication (See the Discussion section). It is noteworthy that while there is mode crosstalk in the transmission link (With unitary crosstalk matrix $C$), the mode sorter can still be intelligently configured to cancel it:

$$\begin{pmatrix} 0 & 0 & 0 & 1 \\ 1 & 0 & 0 & 0 \\ 0 & 1 & 0 & 0 \\ 0 & 0 & 1 & 0 \end{pmatrix} = U_{MZI} S_{LP/OAM} C. \quad (3)$$

We fabricate the mode processor on a standard 220 nm SOI wafer (Fig. 1(e)). The structure of the MZI mesh follows the topology proposed by Clement et al [24]. The TiN heaters on the MZI can be electrically tuned to modify the unitary transmission matrix. The size of the 2×2 grating array is 80×80 $\mu m^2$. The chip is optically and electrically well packaged except for the 2×2 grating array, which will emit/receive the optical modes to/from the free space (Fig. 1(d)).

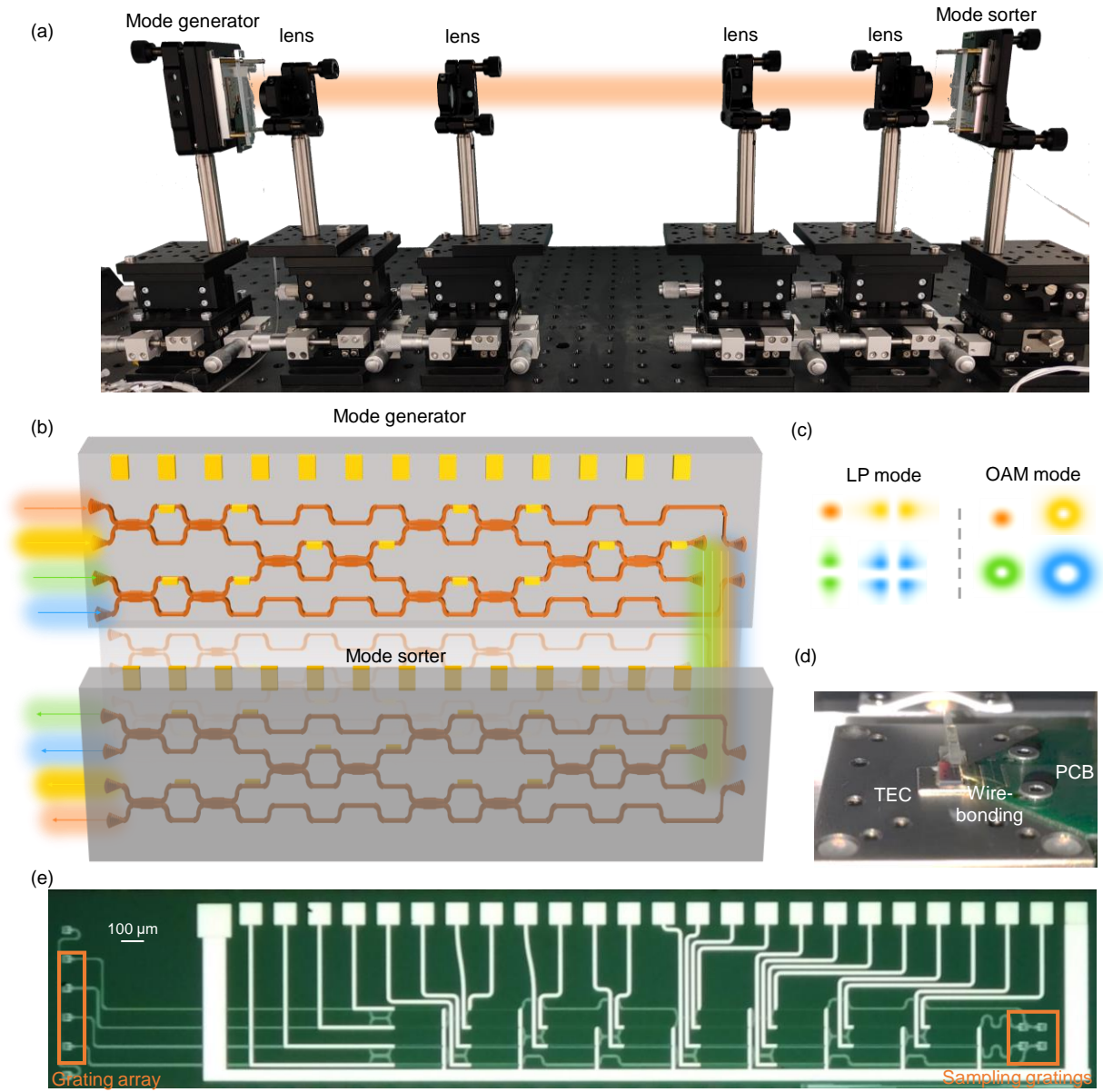

Fig. 1 Working principle and chip fabrication of the intelligent mode processor. (a) The chip-to-chip MDM system setup. (b) The mode processor consists of four aligned single-mode gratings, a 4×4 optical MZI unitary matrix and a 2×2 grating antenna transmitter/receiver. The mode generator generates orthogonal optical modes (LP modes or OAM modes) through customized training. These modes carry different information and are mixed during transmission. The mode sorter is trained to separate these modes and transmit them to the desired output port. (c) The mode patterns of LP modes and OAM modes. (d) The optical and electrical package of the chip. (e) The microscope image of the fabricated chip.

**Training of the optical mode processor**

To train a mode generator, the most efficient method is to train it as a mode sorter because it is easier to generate LP and OAM modes by spatial light modulator (SLM) than to examine them. Fig. 2(a) shows the experimental setup for training the mode generator. The output of a 1560-nm continuous wave laser is expanded and collimated from a single-mode optical fiber into the free space by a fiber collimator. The collimated light is then modulated by an SLM, where the phase distribution of LP and OAM modes is loaded. The phase patterns on the SLM and the corresponding far-field distributions of the optical modes generated by the SLM are shown in Fig. 2(b) (See S2 in the Supplementary material for the validation of the OAM modes). After the spatial Fourier transform of a lens with focal length of 50 mm, the diameter of the collimated light will match the size of the 2×2 grating array, thus ensuring a satisfactory coupling efficiency. The chip is controlled by a field programmable gate array (FPGA) module and intelligent algorithm. A lot of algorithms can be employed to train the MZI mesh, like particle swarm optimization, genetic algorithm and gradient decent algorithm (GDA) [25-27]. We adopt GDA here because it will converge faster in a small model. The loss function of the training is defined as

$$L = -\min([ER_1, ER_2, ER_3, ER_4]), ER_k = P_k/(P_{total} - P_k), \quad (4)$$

where $P_k$ and $P_{total}$ are the output power of the desired output port and the total output power of all output ports respectively, when the *k*-th optical mode is input. $ER_k$ represents the extinction ratio of the *k*-th mode. The process of the gradient decent algorithm is as follows.

(I) Randomly initialize the voltages applied to the heaters;
(II) Increase the voltage by 0.01 V for each heater and evaluate the loss function $L(U+0.01)$ respectively;
(III) Decrease the voltage by 0.01 V for each heater and evaluate the loss function $L(U-0.01)$ respectively;
(IV) Estimate the gradient of the loss function by

$$G = \frac{L(U+0.01) - L(U-0.01)}{0.02}. \quad (5)$$

(V) Update the voltages using the Adam algorithm, which is a fast-converging gradient decent algorithm [28]

$$U(iter+1) = U(iter) - \alpha(v_{iter}/1-\beta_1^{iter})/\sqrt{s_{iter}/(1-\beta_2^{iter})+\epsilon},$$
$$v_{iter} = \beta_1 v_{iter-1} + (1-\beta_1)G, \quad (6)$$
$$s_{iter} = \beta_2 s_{iter-1} + (1-\beta_2)G^2,$$

where *iter* is the current iteration, $\alpha$ is the learning rate which is set to be 0.05 during training, $\beta_1$, $\beta_2$ and $\varepsilon$ are set to be 0.9, 0.999 and $10^{-8}$, respectively. The initial values of $v_{iter}$ and $s_{iter}$ are zero.

(VI) Repeat steps (II-V) until the loss function converges and the best voltages are stored.

Fig. 2(c) shows the evolution of the minimum *ER* during the training of the mode routing in Fig. 2(a). At training epoch of 155, the minimum *ER* is more than 15 dB. Fig. 2(d) shows three mode transmission matrices at the different stages of training, indicating the effectiveness of the training. Using the method described above, we obtain a set of voltages that can generate four OAM modes. (See S2 in the Supplementary material for more experimental results on the sorting of SLM-generated LP and OAM modes).

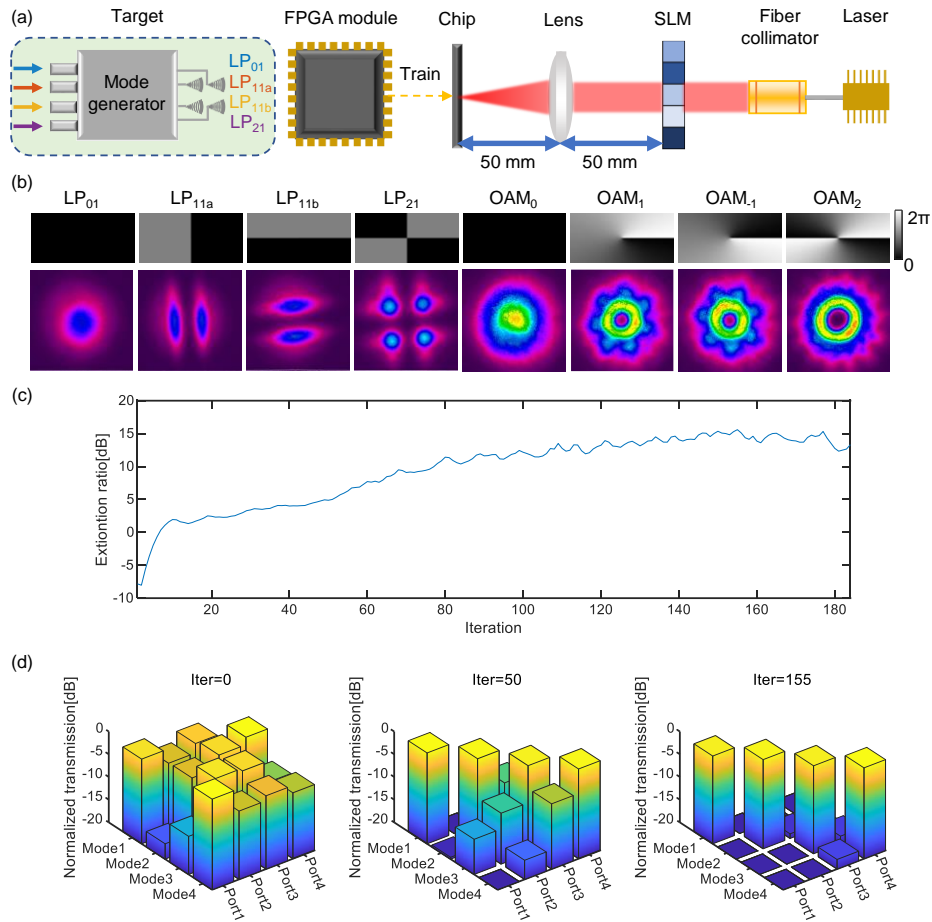

Fig. 2 Training the generator chip. (a) Experimental setup for training the mode generator. The phase distribution of the four LP modes is loaded onto the spatial light modulator (SLM), and a lens is used to match the size of the generated LP modes and the grating antenna array. (b) The far-field distribution of the optical modes generated by the SLM with corresponding phase distribution. (c) The evolution of the lowest extinction ratio among the four LP modes during training. (d) The bar chart of the routing state during training.

Fig. 3(a) shows an experimental setup to study the generated modes with the chip. The mode is first demodulated by the SLM and a CCD is placed behind the SLM to capture the far-field interference pattern. When no phase pattern is loaded on the SLM, the far-field pattern will reflect the mode type. As shown in Fig. 3(b), the far-field pattern of the LP modes is a two-dimensional spot array and different modes have different center position offsets. Compared with the $LP_{01}$ mode, the $LP_{11a}$ and $LP_{11b}$ modes have a horizontal and vertical half-period offset, while the $LP_{21}$ mode has a simultaneous horizontal and vertical half-period offset. For the OAM mode, the $OAM_0$ and $OAM_2$ modes have the same pattern as the $LP_{01}$ and $LP_{21}$ modes because their sampled vectors from the 2×2 grating array are the same. Compared with the $LP_{01}$ mode, the patterns of the $OAM_1$ and $OAM_{-1}$ modes are rotated by 45 degrees, and become blurred since the phase difference of adjacent spots is ±π/2. The experimental results are in good agreement with the simulation (see S3 in the Supplementary material), indicating the successful generation of the designated optical modes. After demodulation by the SLM, the corresponding mode will have a bright spot in the center, while the center of other modes will be dark. In Fig. 3(b), only the diagonal patterns show bright centers, indicating that the generated modes have high purity.

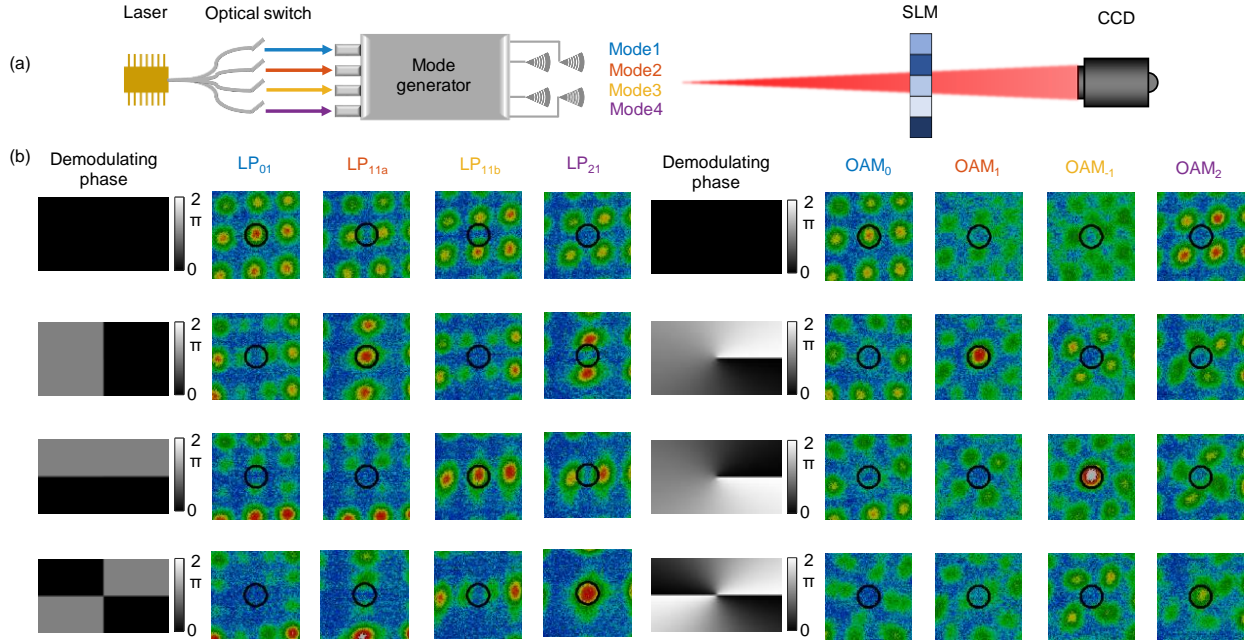

Fig. 3 Verification of chip-generated orthogonal modes. (a) The experimental setup for testing the modes generated by the trained mode generator. The SLM is used to demodulate the four orthogonal modes. A CCD is placed behind the SLM to capture the far-field interference pattern. (b) The far-field interference pattern of the demodulated LP and OAM modes (generated by the mode generator). The interference centers are marked with black circles.

## Demonstration of the MDM system

To demonstrate a real application of the intelligent mode processor, we build a chip-to-chip MDM communication system. As shown in Fig. 4(a), the mode generator and sorter are connected by a lens system. The lens, with a short focal length of 25 mm, is used to effectively collimate the dispersed optical mode so that the alignment of the lens system can be easier. After calibrating of the optical path, we can measure the insertion loss of the eight optical modes generated by the mode generator (see Table 1). Considering the loss of the grating coupler and the MZI mesh (about 25 dB), the extra loss in the 4f coupling is 5~8 dB, which is acceptable in the MDM system.

Table 1. Insertion loss of the chip-to-chip MDM system

| Mode | $LP_{21}$ | $LP_{01}$ | $LP_{11a}$ | $LP_{11b}$ | $OAM_0$ | $OAM_1$ | $OAM_{-1}$ | $OAM_2$ |
|---|---|---|---|---|---|---|---|---|
| Insertion loss (dB) | 29.79 | 32.33 | 32.88 | 31.61 | 30.81 | 32.65 | 33.08 | 31.08 |

The mode sorter is trained using the same rule as described above. Figs. 4(b) and 4(c) show the evolution of the minimum extinction ratio and the three mode transmission matrices at different stages of training. The extinction ratio in the chip-to-chip experiment (18 dB~20 dB) is generally higher than that in the SLM-space-chip experiment (about 15 dB), because the 4f system can more easily satisfy the orthogonality condition of mode sampling. To demonstrate the reconfigurability of our system, three other routing schemes are trained and their final mode transmission matrices are shown in Fig. 4(d), where a mode extinction ratio of more than 18 dB can be achieved. Theoretically, any routing with a high mode extinction ratio can be achieved. (See S4 in the Supplementary material for more experimental results on the performance of the MDM system).

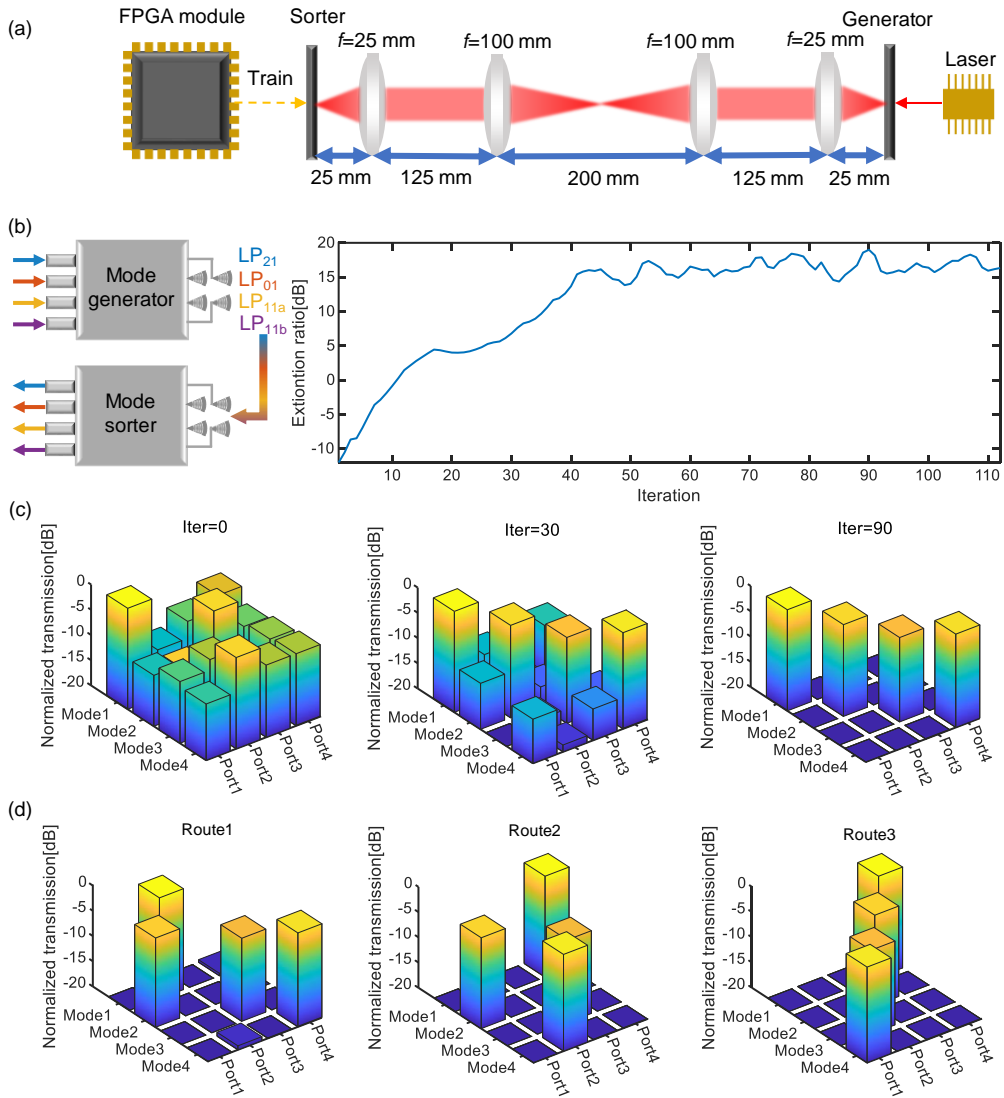

Fig. 4 The training of chip-to-chip MDM communication. (a) Experimental setup of the MDM communication system. A 4f system consisting of four lenses perfectly transmits the generated modes to the grating antenna array of the mode sorter. (b) The evolution of the lowest extinction ratio among four chip-generated LP modes during training. (c) The bar chart of the routing state during training. (d) The bar charts of three different routing schemes after their respective training.

We also test the spectral response of the system. The results are shown in Figs. 5(a) and 5(b) for LP modes and OAM modes, respectively. We can obtain a mode extinction ratio of more than 18 dB at a single wavelength and more than 10 dB at a bandwidth of 10 nm. The bandwidth can be further increased by designing a wavelength-insensitive MZI mesh. To demonstrate its potential in high-speed communications, a 25 Gbit/s pseudorandom binary sequence (PRBS) optical signal is divided equally among the four ports of the mode generator. After training the mode sorter, the undistorted signal is output from the output ports of the mode sorter. On the right side of Figs. 5(a) and 5(b), we show the eye diagrams of the demodulated signals (see S5 in the Supplementary material for the original and the distorted eye diagram). Some degradation of the signal quality (mode2 and mode3 of OAM) is attributed to the imbalance of the insertion loss and the interference of different modes. Overall, the demonstration of the chip-to-chip MDM communication verifies the practicality of our proposed scheme.

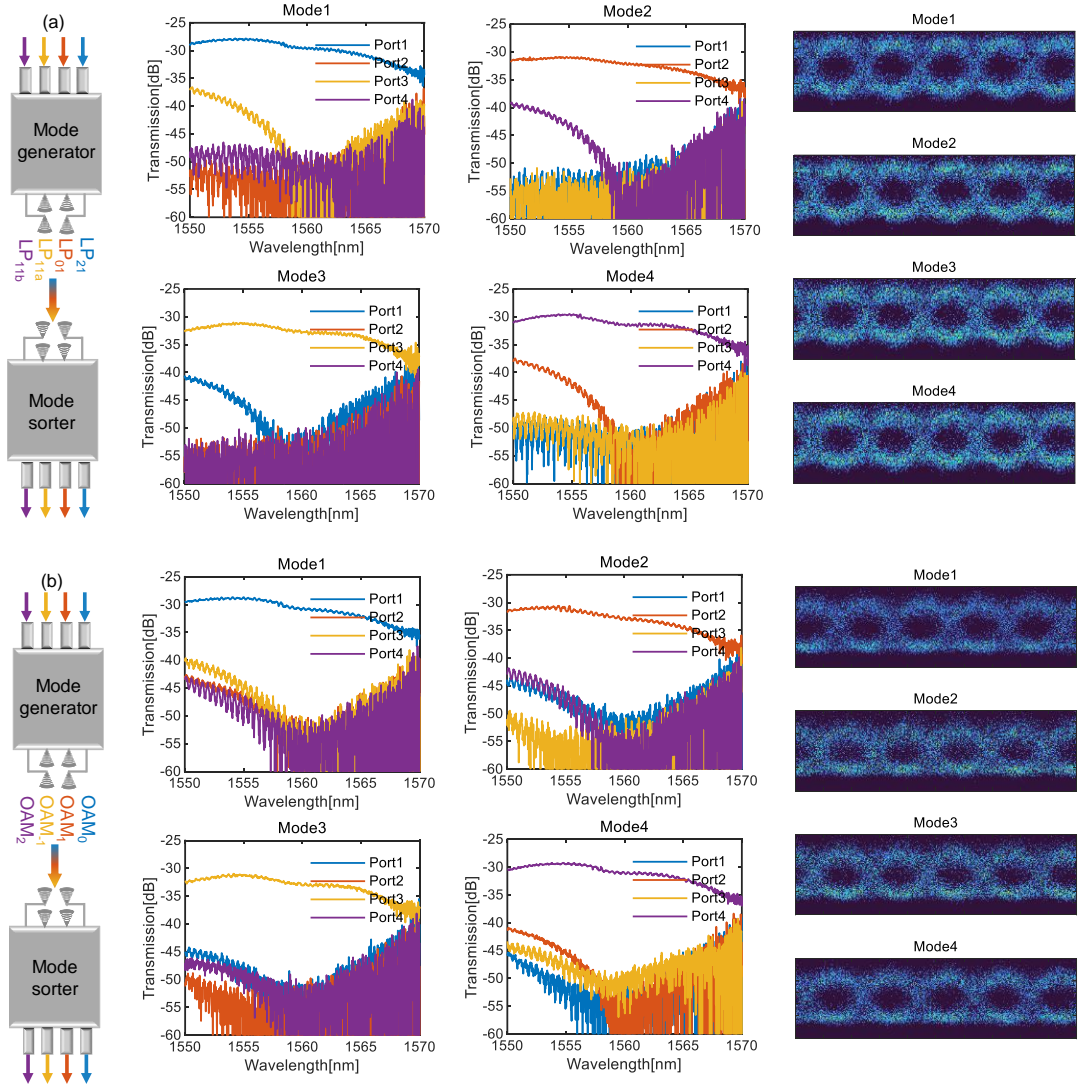

Fig. 5 The optical spectral response and high-speed performance of the proposed optical MDM communication system for (a) LP modes and (b) OAM modes. The insertion loss includes the loss of the two chips and the loss during the transmission. The eye diagrams are captured with a 25 Gbit/s PRBS optical signal simultaneously incident on the four ports of the mode generator.

## Discussion

### Scalability of the optical mode processor

In this section, we will provide a comprehensive illustration of the scalability of our proposed optical mode processor. In Eq. 1, $S_{LP}$ and $S_{OAM}$ correspond to a Hadamard matrix and a discrete Fourier transform (DFT) matrix, respectively. These two types of matrices are typical unitary matrices. We will now show that evenly spaced circular grating couplers can achieve the orthogonal sampling of the higher-order LP and OAM modes. For an $N \times N$ DFT matrix, the element in the $m$-th row and the $n$-th column is $e^{-j2\pi(m-1)(n-1)/N}$. The trajectory of the complex angle of the $n$-th column vector can be seen to rotate from 0 to $2\pi$ for $n-1$ times, which is consistent with the phase distribution of the $OAM_{n-1}$ modes. Moreover, the $OAM_k$ and $OAM_{k-N}$ ($k>0$) will have the same sampled phase distribution when there are $N$ grating couplers. Therefore, some higher-order OAM modes can be replaced by the lower order modes to ensure a more reasonable resource allocation. The same grating coupler distribution can also support LP modes. In this case, the sampled matrix becomes a Hadamard matrix. The $N$ grating couplers can support the modes $LP_{01}$, $LP_{11a}$, $LP_{11b}$, $LP_{21a}$… $LP_{\frac{N}{2}1}$ modes. In Fig. 6, an example of $N=8$ is given, and we annotate each orthogonal mode and its sampled vector. For LP modes, the radius of the sampling circle is crucial because the sampled optical intensity should be the same to ensure orthogonality, while for OAM modes it is not an issue due to their circular symmetry. The scalability of the proposed mode processor makes it very promising for high-capacity communications.

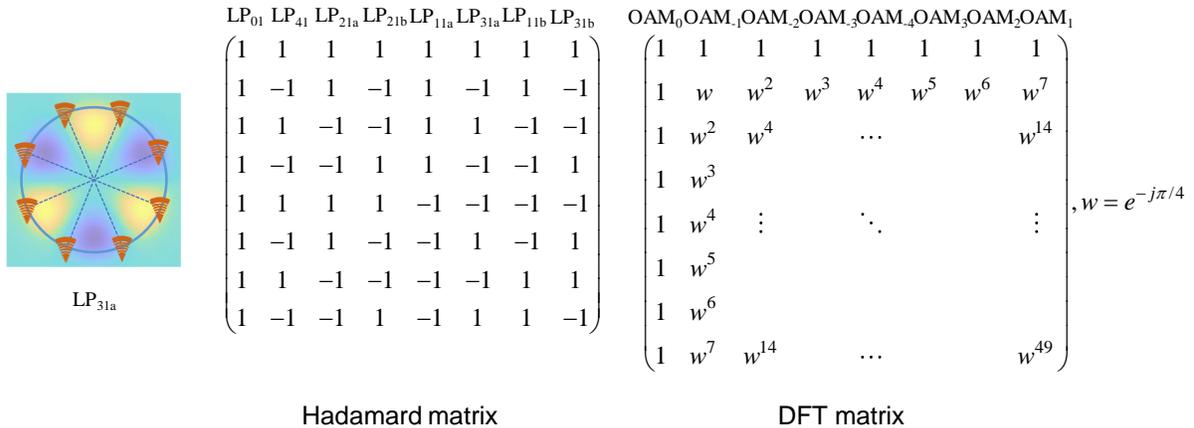

Fig. 6 Orthogonal sampling of LP and OAM modes by eight grating couplers arranged evenly in a circle.

**Potential application in long-distance communication**

In our experiment, the mode generator and the sorter are connected by free space, which is not an ideal scheme for long-distance communication. Multimode fibers and ring-core fibers are the more competitive transmission media for LP and OAM mode communication because the modes and signals in the fibers are less affected by environmental disturbances. Although there still exists mode crosstalk during the fiber transmission, it can be cancelled by intelligently configuring the mode sorter. Considering more complex scenario where temporal crosstalk also occurs, we can adopt the similar mode sorter structure as in [23] to descramble mode-temporal crosstalk. Another key design issue is to match the mode profile of the fiber and the size of the grating antenna. Currently, our grating coupler has a size of 30×50 μm$^2$, while the diameter of the commercial multimode fiber is 62.5 μm. Obviously, the size of the grating coupler is still too large. To solve this problem, the size of the grating coupler can be tailored and demonstrations with moderate performance have been reported [29,30]. In addition, the size of the ring-core fiber should also be carefully designed to match the grating array so that the LP and OAM modes can be randomly switched with the same optical mode generator/sorter. Overall, the proposed architecture is promising for future long-hual fiber communication systems.

**Monolithic integrated silicon photonic multi-dimensional transceiver**

The proposed mode processor has great potential to be monolithically integrated with other multiplexing technology, i.e., WDM (wavelength division multiplexing) and PDM (polarization division multiplexing). Fig. 7 shows a conceptual diagram of the monolithic integrated silicon photonic multi-dimensional transceiver, where the Mach Zehnder modulator (MZM), arrayed waveguide grating (AWG), polarization beam splitter (PBS) and photodetector (PD) can all be integrated with the mode processor on the same silicon chip. There are just a few more considerations: (I) The MZI unitary matrix must be polarization and wavelength-insensitive by designing polarization and wavelength-insensitive 2×2 multimode interferometer (MMI) [31]. (II) The grating coupler antenna can also be specifically designed to be polarization-insensitive [32]. With the state-of-art technology, we can adopt two polarizations, 32 wavelengths with 100 GHz frequency space in the C band, 8 LP/OAM modes and 25 Gbaud 32QAM modulation format [33,34]. In this case, the communication capacity can reach 64 Tbit/s in a single optical fiber.

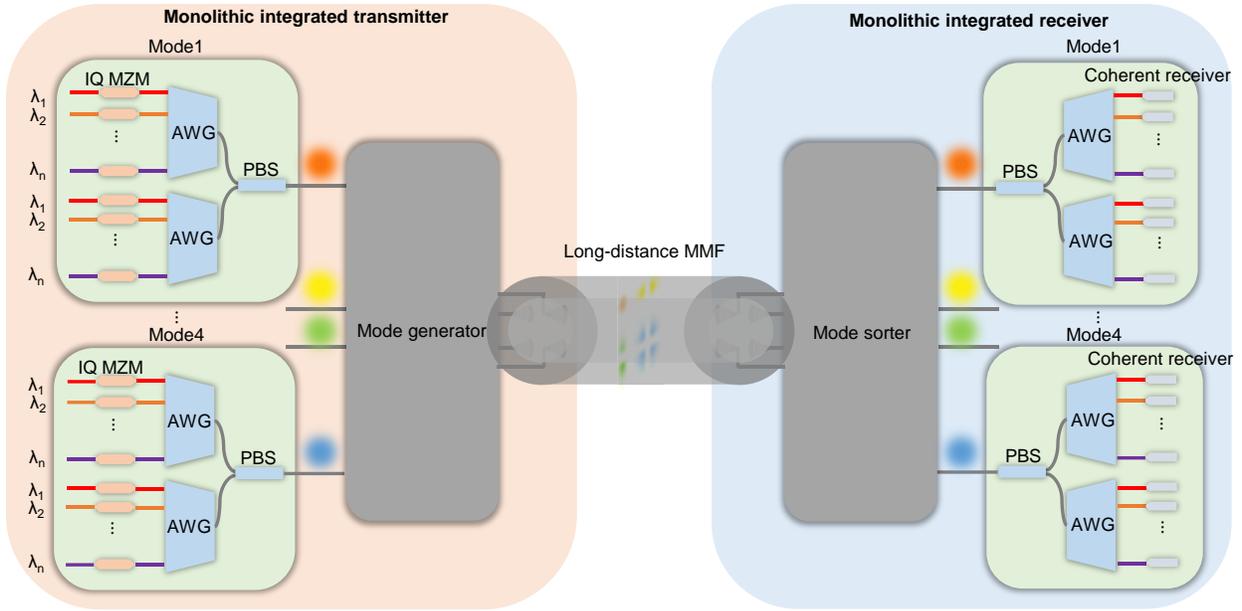

Fig. 7 The conceptual diagram of monolithic integrated silicon photonic multi-dimensional transceiver. MZM, Mach Zehnder modulator. AWG, arrayed waveguide grating. PBS, polarization beam splitter. MMF, multimode fiber.

## Conclusions

In conclusion, we have proposed and demonstrated an intelligent optical MDM system based on the reconfigurable MZI unitary matrix. Theoretically, we showed that the optical mode processor can handle both the LP and OAM modes and support any desired routing scheme. Experimentally, we verified the intelligence of the mode processor and carried out a 25 GHz chip-to-chip communication experiment, demonstrating the capability of multimode communication with our mode processor. Furthermore, we have demonstrated its scalability, potential applications in long-distance communication and compatibility with other multiplexing technologies. It shows significant advantages in universality, intelligence, programmability and resistance to disturbances and fabrication errors. Our work paves the way for future intelligent optical MDM communications.

## Methods

### Details of the experiment

To train the mode generator, the 1560-nm CW laser is first amplified to 23 dBm by an EDFA. After propagating through the SLM (Holoeye HED-6001), the 50 mm focal length lens, and the mode sorter, the optical power output from the chip is measured by a photodetector array electrically connected to the analog-to-digital converter (AD7606). The voltage applied to the heater is provided by the digital-to-analog converter (LTC2688). Both the AD7606 and the LTC2688 are controlled by a field programmable gate array chip (7K325T). The entire experimental system was controlled by a personal computer through serial ports. The mode generator and mode sorter are thermally stabilized by a thermoelectric cooler (TEC). We use the amplified spontaneous emission from the EDFA as the input source, and an optical spectrum analyzer (Yokokawa AQ6319) captures the spectrum of the input source and the output from the chip. The difference between the two measurements is taken as the transmission spectrum of the system. In the high-speed communication experiment, the 1560-nm CW laser is first injected into an optical modulator and modulated by the 25G PRBS code generated by a bit pattern generator (BPG). The modulated optical signal is amplified and split by a 1×4 power splitter. The four channels operate simultaneously and the oscilloscope records the eye diagrams of the signals received by a photodetector (Finisar).

**The characterization of fabricated chip**

The fabricated grating coupler has a diffraction angle of 12°. Therefore, the two chips are titled in the chip-to-chip communication experiment to ensure optical alignment. At the working wavelength of 1560 nm, the insertion loss of a single grating coupler is about 5 dB. The insertion loss of the 4×4 MZI unitary matrix is about 2.5 dB, which is mainly due to the loss of the 2×2 multimode interferometer (MMI). The total chip loss is about 12.5 dB. The π phase shift power consumption of a single MZI is about 18.4 mW.

## Declaration

### Ethics approval and consent to participate

There is no ethics issue for this paper.

### Consent for publication

All authors agreed to publish this paper.

### Availability of data and materials

The datasets used and/or analyzed in the current study are available from the corresponding author upon reasonable request.

### Competing interests

The authors declare that they have no competing interests.

### Funding

National Natural Science Foundation of China (62275088, 62075075, U21A20511); Innovation Project of Optics Valley Laboratory (Grant No. OVL2021BG001); Research Grants Council, University Grants Committee of Hong Kong SAR under Grant PolyU15301022.

### Authors' contributions

HZ, JD and BW conceived the idea. BW and WZ designed and fabricated the chip. BW and WZ designed and performed the experiments. HZ, JD, BW and WZ discussed and analyzed data. BW prepared the manuscript. HZ, JD, DH and PW revised the paper and XZ supervised the project. All authors contributed to the writing of the manuscript.

**Acknowledgements**

Not applicable.

**Authors' information**

Not applicable.

# Supplementary materials for "Chip-to-chip optical multimode communication with universal mode processors"

Bo Wu[1,†], Wenkai Zhang[1,†], Hailong Zhou[1,*], Jianji Dong[1,*] Dongmei Huang[2, 3], P. K. A. Wai[4] and Xinliang Zhang[1]

[1] *Wuhan National Laboratory for Optoelectronics, School of Optical and Electronic Information, Huazhong University of Science and Technology, Wuhan 430074, China*

[2] *The Hong Kong Polytechnic University Shenzhen Research Institute, Shenzhen 518057, China*

[3] *Photonics Research Institute, Department of Electrical Engineering, The Hong Kong Polytechnic University, Hong Kong, 999077, China*

[4] *Department of Physics, Hong Kong Baptist University, Kowloon Tong, Hong Kong, 999077, China*

[†]*These authors contributed equally to this work*

[*]Corresponding author: hailongzhou@hust.edu.cn; jjdong@hust.edu.cn;

## S1. The design of optical MZI unitary matrix

The basic unit of a 4×4 optical unitary matrix is a 2×2 MZI, which is enclosed by the blue dashed box in Fig. (S1). Its transmission matrix can be written as [1]

$$M = T_{PS2}T_{MMI}T_{PS1}T_{MMI} = \begin{pmatrix} e^{i\varphi_2} & 0 \\ 0 & 1 \end{pmatrix}\begin{pmatrix} \sqrt{2}/2 & \sqrt{2}i/2 \\ \sqrt{2}i/2 & \sqrt{2}/2 \end{pmatrix}\begin{pmatrix} e^{i\varphi_1} & 0 \\ 0 & 1 \end{pmatrix}\begin{pmatrix} \sqrt{2}/2 & \sqrt{2}i/2 \\ \sqrt{2}i/2 & \sqrt{2}/2 \end{pmatrix}. \tag{S1}$$

This matrix and its inverse can achieve the column and row elimination of any element in a 2×2 matrix respectively. For a 4×4 unitary matrix $U$, eliminate its elements following the sequence: $U_{41}$-$U_{31}$-$U_{42}$-$U_{43}$-$U_{32}$-$U_{21}$ and the eliminating method: column-row-row-column-column-column. The eliminating procedure can be written as

$$T_{3,4}^{-1}T_{2,3}^{-1}UT_{1,2}T_{3,4}T_{2,3}T_{1,2}, \tag{S2}$$

where the $T_{k,k+1}$ is the MZI eliminating matrix (replacing the elements $(k, k)$, $(k, k+1)$, $(k+1, k)$ and $(k+1, k+1)$ of an eye matrix with a 2×2 MZI matrix). According to the property of unitary transformation, the resulting matrix becomes a diagonal unitary matrix after elimination. Therefore, the 4×4 optical unitary matrix can be written as

$$U = T_{2,3}T_{3,4}DT_{1,2}^{-1}T_{2,3}^{-1}T_{3,4}^{-1}T_{1,2}^{-1} = T_{2,3}T_{3,4}T_{1,2}T_{2,3}T_{3,4}T_{1,2}D', \tag{S3}$$

where the diagonal unitary matrix $D$ can be transferred to the right-most side of the matrices [1]. The decomposition of the unitary matrix can be transformed to the physical architecture in Fig. S1, where the diagonal matrix $D'$ is not necessary in our application and thus deleted here. Actually, there are additional phase shifters in the fabricated chip to compensate the phase difference.

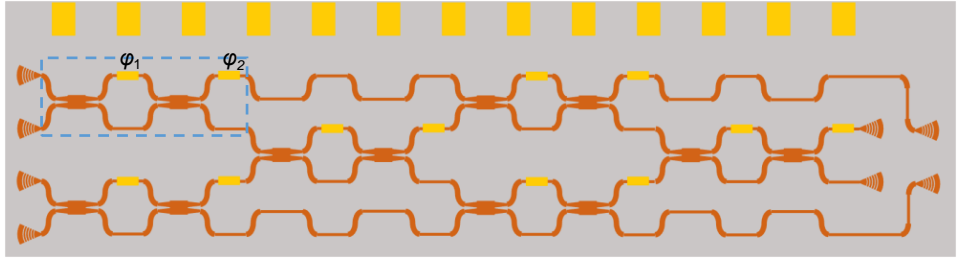

Fig. S1 The architecture of a 4×4 unitary MZI matrix.

## S2. The experimental results for OAM/LP mode sorting

To investigate the SLM-generated OAM modes, we use the coordinate transformation method proposed in [2], in which we add a phase distribution of a cylindrical lens to the SLM. In the far field, we can capture the patterns of different OAM modes (see Fig. S2), which are consistent with the results of [2]. In the following, we show the rest of the experimental results of training the optical mode generator, which includes two routing schemes of LP modes and three routing schemes of OAM modes (see Fig. S3). All the results show large mode extinction ratios.

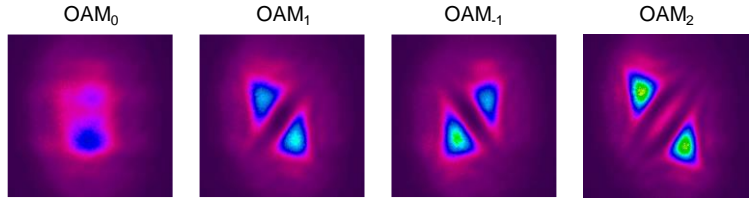

Fig. S2 The interference pattern of SLM-generated OAM modes by the coordinate transformation with cylindrical lens.

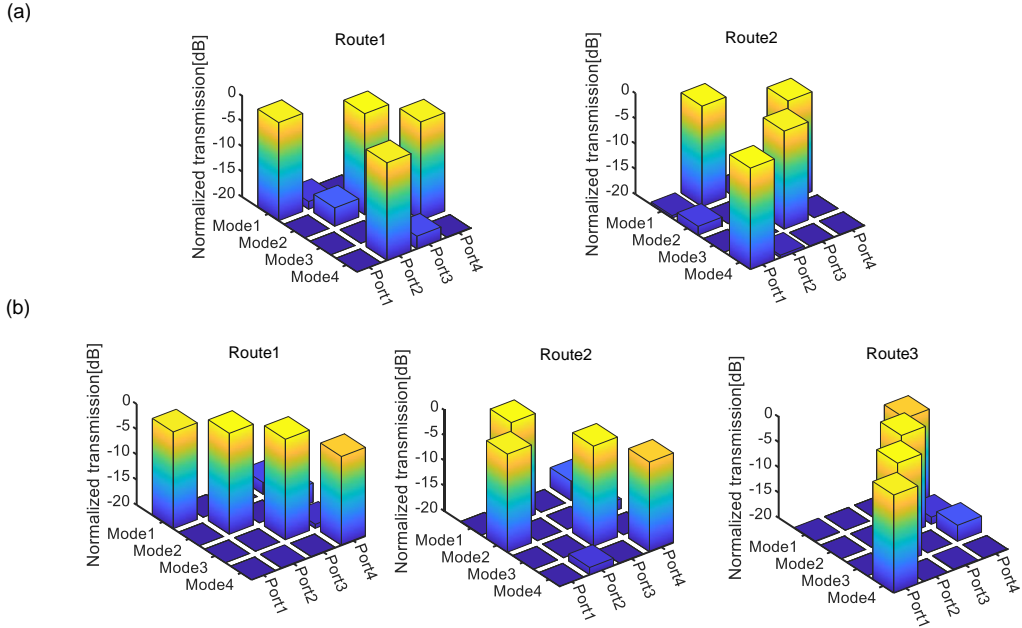

Fig. S3 The bar charts of more routing schemes of SLM-generated (a) LP modes and (b) OAM modes.

## S3. The simulation of far-field patterns for OAM/LP mode generation

To further explain the far-field pattern in Fig. 3(b), we show a numerical simulation to compare with the experimental observation. In the simulation, we generate four Gaussian optical spots with different phase to simulate the actual emission of four grating couplers

$$E(x,y) = e^{i\varphi_1 - \left[(x-d/2)^2 + (y-d/2)^2\right]/w^2} + e^{i\varphi_2 - \left[(x+d/2)^2 + (y-d/2)^2\right]/w^2} + e^{i\varphi_3 - \left[(x-d/2)^2 + (y+d/2)^2\right]/w^2} + e^{i\varphi_4 - \left[(x+d/2)^2 + (y+d/2)^2\right]/w^2}, \quad (S4)$$

where in the simulation $d$=80 μm and $w$=10 μm. We take $[\varphi_1, \varphi_2, \varphi_3, \varphi_4]$ to be $[0, 0, 0, 0]$, $[0, \pi, \pi, 0]$, $[0, 0, \pi, \pi]$ and $[0, \pi, 0, \pi]$ to represent quasi-LP$_{01}$, LP$_{11a}$, LP$_{11b}$ and LP$_{21}$ modes. According to the Fresnel diffractive formula,

$$E(x,y) = \frac{e^{ikz}}{i\lambda z} \iint E(x_1, y_1) e^{\frac{ik}{2z}\left[(x-x_1)^2 + (y-y_1)^2\right]} dx_1 dy_1, \quad (S5)$$

where $k$ is the wave vector and $\lambda$ is 1560 nm, we can simulate the optical spot for the propagation distance $z$=1 mm (see Fig. S4). Similarly, let $[\varphi_1, \varphi_2, \varphi_3, \varphi_4]$ be $[0, 0, 0, 0]$, $[0, 0.5\pi, \pi, 1.5\pi]$, $[0, 1.5\pi, \pi, 0.5\pi]$ and $[0, \pi, 0, \pi]$, we can simulate the far-field interference pattern of the quasi-OAM modes. The simulation results are consistent with the experimental results shown in the main text, indicating that the correct modes are generated by the mode generator.

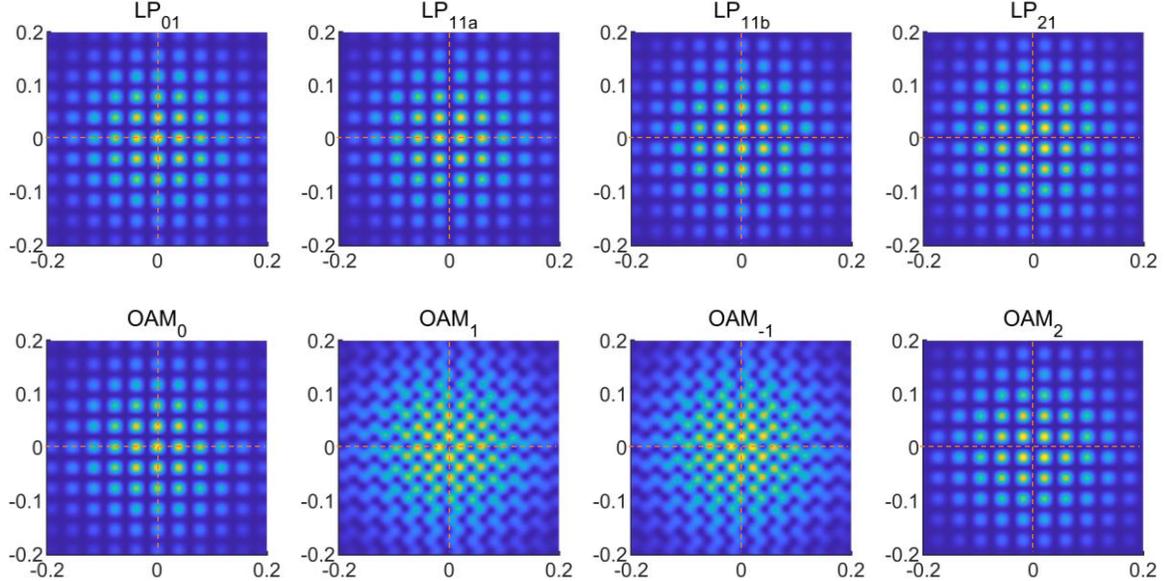

Fig. S4 The simulated far-field interfering pattern of the quasi-LP and OAM modes.

## S4. The experimental results for optical multimode communication

In this subsection, we show further experimental results of the chip-to-chip MDM communication in Fig. S5 to validate the reconfigurability of the mode processor.

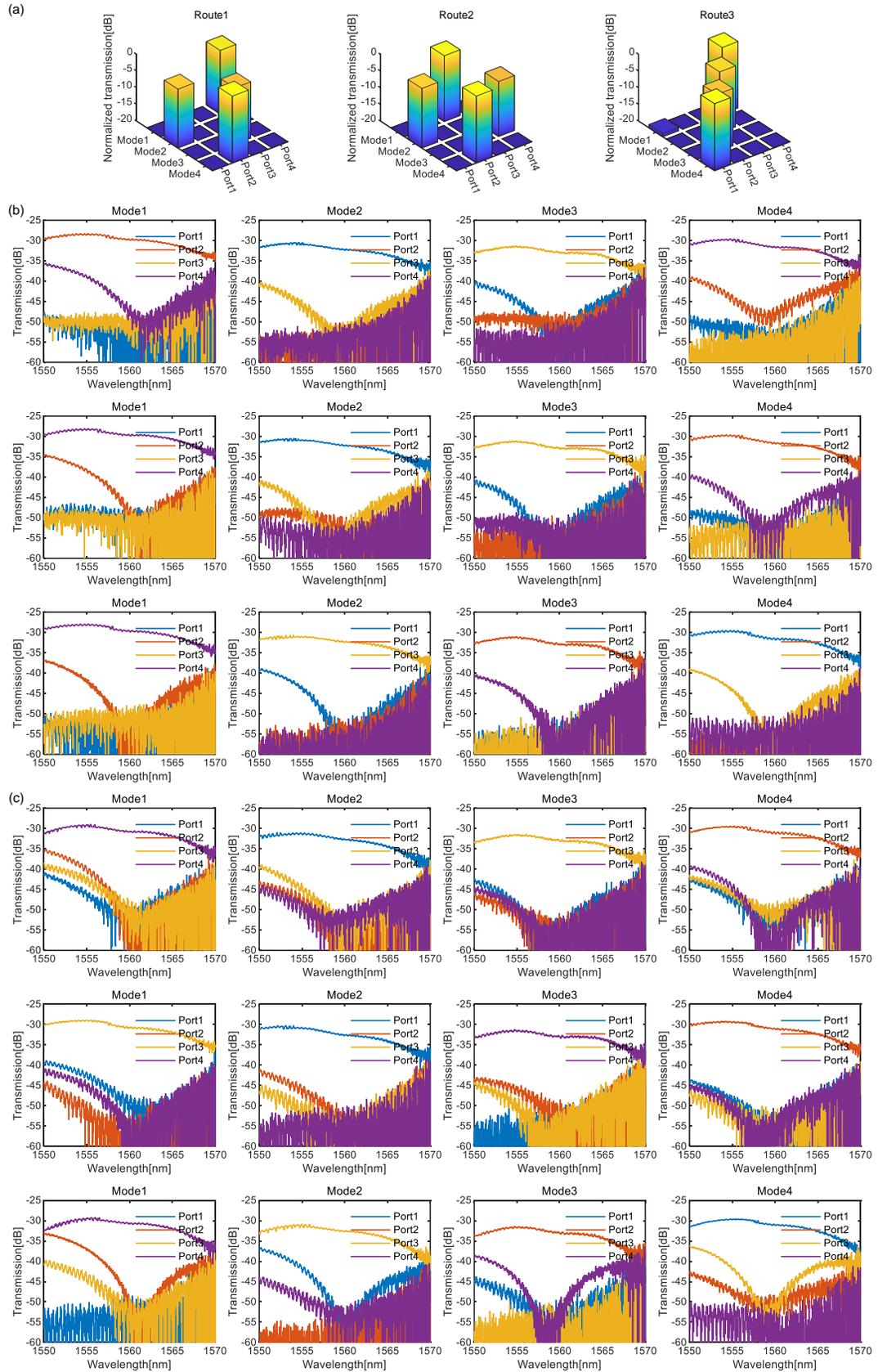

Fig. S5 Additional results of the MDM communication experiment. (a) Three more routing schemes of OAM modes. (b) The optical spectral response of the three routing schemes in Fig. 4(d). (c) The optical spectral response of the three routing schemes in (a).

## S5. The eye diagram of the input optical signal and untrained output signal

For comparison, the eye diagrams of the input optical signal and the untrained output of the mode sorter are shown below. Owing to the imperfection of the original electrical signal generated by the BPG, the quality of the eye diagram of the input signal is only moderate. However, the eye diagrams in the main text do not degrade significantly, which means that the MDM system preserves the quality of the optical signals well.

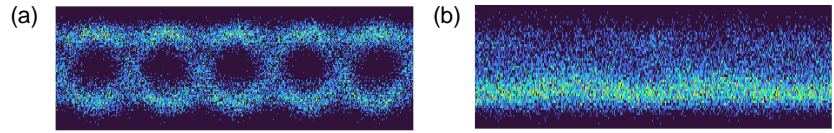

Fig. S6 (a) The eye diagram of the input optical signal. (b) The eye diagram of the untrained output of the mode sorter.